\documentclass{PoS}

\title{The Muon g-2 Experiment Overview and Status}

\ShortTitle{The Muon g-2 Experiment Overview and Status}

\author{\speaker{J. L. Holzbauer}\\
        University of Mississippi\\
        E-mail: \email{jholzbau@fnal.gov}}

\abstract{The Muon g-2 experiment at Fermilab will measure the anomalous magnetic moment of the muon to a precision of 140 parts per billion, which is a factor of four improvement over the previous E821 measurement at Brookhaven. The experiment will also extend the search for the muon electric dipole moment (EDM) by approximately two orders of magnitude. Both of these measurements are made by combining a precise measurement of the 1.45T storage ring magnetic field with an analysis of the modulation of the decay rate of the higher-energy positrons from the (anti-)muon decays recorded by 24 calorimeters and 3 straw tracking detectors. The current status of the experiment as well as results from the initial beam delivery and commissioning run in the summer of 2017 will be discussed.}

\FullConference{The 19th International Workshop on Neutrinos from Accelerators-NUFACT2017\\
		25-30 September, 2017\\
		Uppsala University, Uppsala, Sweden}

\begin{document}

\section{Overview}
The Muon g-2 experiment at Fermilab is designed to measure the anomalous magnetic moment of the muon to high precision~\cite{tdr}.  It is an improved version of an experiment at BNL called E821, which measured the value to be approximately three standard deviations away from the theoretical value~\cite{e821}.  Such an interesting possibility of new physics in a precision experiment makes the new measurement extremely important.  The experiment is designed to make the measurement more precisely and potentially indicate whether or not there is really a sign of new physics in this sector.

In Dirac theory, the g factor relates the spin and the magnetic moment of the muon and is exactly two~\cite{dirac}.  However, higher order effects cause the value to be slightly different, and this deviation from two is known as the anomaly, namely $a_{\mu}=$(g-2)/2.  If the anomaly varies from the standard model expectation, it can indicate new physics, such as a heavy W in certain SUSY models~\cite{snowmass}.  This anomaly is the value the Muon g-2 experiment is designed to measure, and also contributes to its name.

The standard model value of $a_{\mu}$ is estimated separately by theorists.  It has several parts:  quantum electrodynamics (QED)~\cite{smqed}, electroweak (EW)~\cite{smew}, hadronic vacuum polarization (HVP)~\cite{smhpv1,smhpv2} and hadronic light-by-light (HLbL)~\cite{smhlbl}.  As can be seen in Table~\ref{SM}, the QED and EW parts are known very well.  The others, being hadronic terms, are less well known and are estimated using various techniques, including data from other experiments (such as e+e- to hadrons), and lattice calculations.

The E821~\cite{e821} measurement at BNL, corrected for updated constants~\cite{codata} and given in~\cite{tdr}, is $a_{\mu}^{E821} = 116 592 089 \pm 63 \times 10^{-11}$ (54~ppm).  The uncertainty is slightly larger than the standard model calculation uncertainty.  It is expected that the current experiment will be able to measure the value with an uncertainty approximately four times lower, 0.14ppm.  If this happens and the central value remains the same, the deviation of $a_{\mu}$ from the standard model expectation would be at least 5 sigma.  Expected improvements to the understanding of the standard model value would give a deviation of 8 sigma under these same assumptions~\cite{tdr}.

\begin{table}[h]
\caption{\label{SM} Standard model components of the anomaly, taken directly from~\cite{tdr}.  Two values are shown for HVP to reflect two recent estimates.  The terms lo and ho indicate lower order and higher order, respectively.  Other terms are defined in the text.}
\begin{center}
\begin{tabular}{lr}
\hline
&Values in $10^{-11}$ units\\
\hline
QED ($\gamma$ + $l$) & 116584718.951 $\pm$ 0.009 $\pm$ 0.019 $\pm$ 0.007 $\pm$ 0.077\\
HVP(lo)~\cite{smhpv1} & 6923 $\pm$ 42\\
HVP(lo)~\cite{smhpv2} & 6949 $\pm$ 43\\

HVP(ho)~\cite{smhpv2} & -98.4 $\pm$ 0.7\\
HLbL         & 105 $\pm$ 26\\
EW           & 153.6 $\pm$ 1.0\\
\hline
Total SM~\cite{smhpv1} & $116591802 \pm 42_{\rm{H-LO}} \pm 26_{\rm{H-HO}} \pm 2_{\rm{other}}$ ( $\pm 49_{\rm{tot}}$)\\
Total SM~\cite{smhpv2} & $116591828 \pm 43_{\rm{H-LO}} \pm 26_{\rm{H-HO}} \pm 2_{\rm{other}}$ ( $\pm 50_{\rm{tot}}$)\\
\hline
\end{tabular}
\end{center}
\end{table}

\section{The Muon g-2 Experiment}
Although the Muon g-2 experiment is located at Fermilab, the similarities in design to E821 suggested the reuse of several parts from the BNL experiment.  Thus, it was decided to move various components across the country, most notably the 15 ton cryostat ring, which was moved by barge and truck, with the requirement that the superconducting coils not flex by more than 3mm.  In the summer of 2015 the dipole magnet and cryo-system were assembled at Fermilab and achieved the 1.45 Tesla field, indicating a successful transport.

The muons for the storage ring are produced by the Fermilab accelerator complex.  The source protons are reused from the Tevatron operations, and they collide with an Inconel target to produce pions.  These pions enter a long delivery ring where they decay to muons and neutrinos before entering the Muon g-2 storage ring as a polarized positively charged muon beam.  The extra time between the target and the storage ring helps to reduce beam contamination.  The new experiment expects a factor of 20 increase in muon statistics over the BNL experiment and which will reduce the statistical error to 0.1~ppm.

In brief, the experiment consists of a cryo-system surrounding C-shaped dipole magnets, which contain rectangular vacuum chambers with electrostatic quadrupole plates and kicker plates inside, depending on the region of the ring.  The magnets and quadrupole plates work in tandem to keep the muons in the proper location after the kickers move the beam into a stable circular orbit.  Other components include an inflector magnet to reduce the field as the beam enters the ring, collimators to control the beam, a beam monitoring system, and a trolley that rides around inside of the ring under vacuum to measure the magnetic field with NMR probes.  Detector systems include both straw trackers and lead crystal calorimeters with multiple readout channels.  The dipole magnet system is C-shaped to allow less material between the calorimeters and the muon decay points.

Most major components like the vacuum chambers for the storage ring, dipole magnets, and electro-static quadrupole plates were reused from the E821 experiment.  However, these components were disassembled and reassembled, including re-shimming the magnet to create a constant field and realigning the quadrupole plates.  Vacuum chambers were modified to include new systems inside the storage ring.  Many smaller components for these various sub-systems are new, and in some cases the assembly procedures are also different than those from E821.  Other changes include improved calorimeter resolution and calibration system, and the presence of a straw tracking system, which is completely new for the current experiment.  The kicker plates were redesigned, as were the collimators.  Additionally, a thinner quadrupole plate was added near the region of beam entry, to reduce muon losses before the beam is kicked into orbit by the kickers, which are located 90 degrees down the ring from the beam entry position.

\section{The Measurement Procedure}
There are two quantities that compose the measurement of $a_{\mu}$, namely $\omega_a$ and $\omega_p$. The value $\omega_a$ is the precession frequency of the muon with respect to the momentum, which is measured via high energy decay positrons, and $\omega_p$ is the magnetic field (B) normalized to the proton lamour frequency.  The spin and cyclotron frequencies can be combined to give $a_{\mu}$ in terms of $\omega_a$, B, and the charge over mass ratio.  This can then be rewritten in terms of $\omega_a$/$\omega_p$ and the muon, proton magnetic moment ratio from hyperfine splitting, which is obtained elsewhere.  Additionally, if one were to write out the full form of $\omega_a$, one would notice an electric field term.  This term will be zero with the appropriate choice of muon momenta, 3.09 GeV/c, known colloquially as the magic momentum, and corresponding to a magic radius (which sets the radius of the storage ring).  Of course, not all muons will have this exact momentum, and this is accounted for as an analysis uncertainty.  This experiment technique also makes alignment efforts particularly important.

The value of $\omega_p$ is obtained from measurements of the magnetic field, which is designed to be as constant as possible.  There is a device called a trolley which has NMR probes mounted on it and traverses the storage ring where the beam is located (when the beam is off).  A second larger trolley was also used to make measurements before the installation of the storage ring vacuum chambers, to have access to higher order field modes.  Additionally, there are NMR probes permanently fixed to the tops and bottoms of the storage ring vacuum chambers.  These can run when the beam is on or off and are used to relate the NMR trolley measurement to a beam-on measurement.  Finally, absolute measurements are taken in a particular region of the ring using specialized probes, and the relative NMR measurements are related to this.

The quantity $\omega_a$ is obtained from high energy positron data.  Calorimeters in 24 stations around the ring record energy deposits from these particles and are used to make the so-called wiggle plot, which shows a varying number of events with time in an oscillatory pattern (in addition to the normal decrease due to a finite number of particles decaying).  Additional information from tracker systems and others help to reduce lost muon contamination and estimate various uncertainties on the measurement.  

\section{2017 Commissioning Run}
\begin{figure}
  \begin{center}
    \includegraphics[width=.6\textwidth]{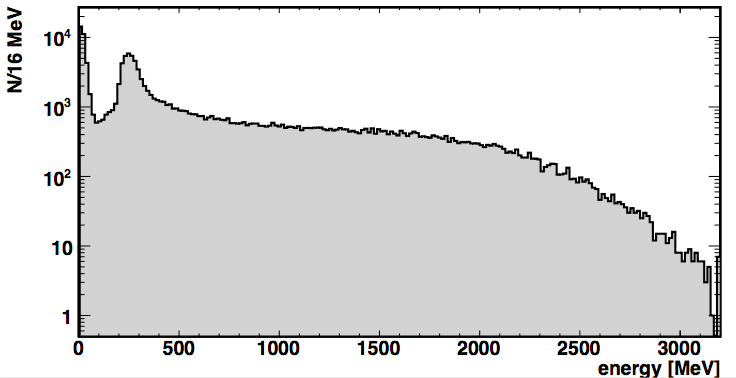}
  \end{center}
  \caption{Energy distribution from June 2017 data recorded in the calorimeter. The peaks are from protons and lost muons.}
  \label{calo}
\end{figure}

\begin{figure}
  \begin{center}
    \includegraphics[width=.6\textwidth]{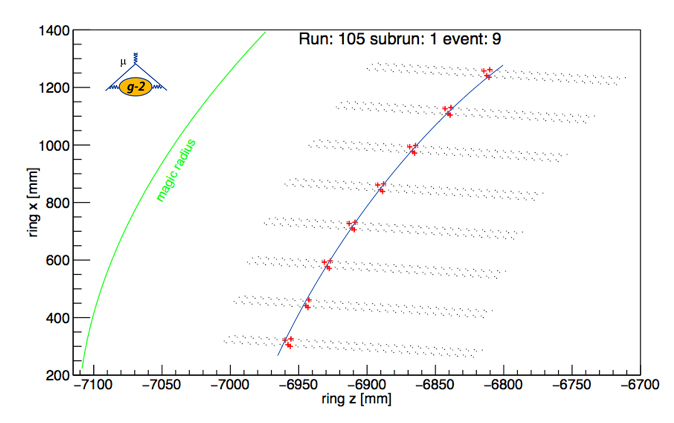}
  \end{center}
  \caption{One of the first tracks recorded by the tracker showing the hits from a single charged particle (likely a proton) through the straw trackers and the (wire)-track fit and the magic radius.}
  \label{track}
\end{figure}

\begin{figure}
  \begin{center}
    \includegraphics[width=.6\textwidth]{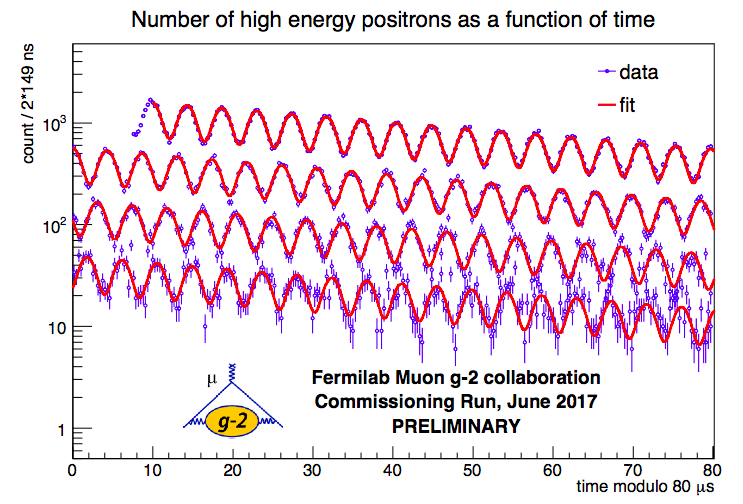}
  \end{center}
  \caption{Distribution of position counts from two weeks of data accumulated in June 2017.}
  \label{wiggle}
\end{figure}

In the summer of 2017, the Muon g-2 experiment had a short commissioning run to test the systems and procedures, during which the accelerator complex ran with a reduced rate relative to the physics run plan.  On May 23, the first particles were delivered to the ring and beam splash was seen in the calorimeters.  The run ended July 7.  Over the course of the run, the experiment was able to achieve particles circulating through the full storage ring and demonstrated the operations of all systems.  In Figure~\ref{calo} we see a distribution of energies detected in the calorimeters, with peaks corresponding to lost muons and protons.  In Figure~\ref{track}, a reconstructed track is shown in two spatial dimensions, compared to the magic radius location, of a positron that traversed several straws, again demonstrating an operational system.  Finally, in Figure~\ref{wiggle}, the so-called wiggle plot is shown for the commissioning data, along with a preliminary analysis fit.  This distribution was created from two weeks of data taken in June of 2017 and contains approximately 700,000 positrons.  The statistics correspond to results similar to that of the older CERN-II experiment which preceded E821~\cite{cern}.  

Overall, the run was quite successful.  The next run, which will be used for physics data analyses, will take place primarily in 2018.  It will be preceded by a second commissioning phase that began in November of 2017 as systems were restarted.

\section{Summary and Plans}
The Muon g-2 experiment is entering an exciting data taking phase.  The machine is assembled and operational, and capable of producing the various quantities needed to make the intended measurement.  It is expected that the new experiment will improve the uncertainty on the measured value versus the previous measurement by a factor of four.  With the data taken in 2018, the Muon g-2 experiment has the potential to finally resolve the tension of the previous result and the standard model calculation. 


\bibliographystyle{iop-num}
\bibliography{holzbauer}

\providecommand{\newblock}{}
\begin{thebibliography}{10}
\expandafter\ifx\csname url\endcsname\relax
  \def\url#1{{\tt #1}}\fi
\expandafter\ifx\csname urlprefix\endcsname\relax\def\urlprefix{URL }\fi
\providecommand{\eprint}[2][]{\url{#2}}

\bibitem{tdr}
Grange J {\em et~al.\/} (Muon g-2) 2015  (\textit{Preprint}
  \eprint{1501.06858v1})

\bibitem{e821}
Bennett G {\em et~al.\/} (Muon G-2 Collaboration) 2006 {\em Phys.Rev.\/} {\bf
  D73} 072003 (\textit{Preprint} \eprint{hep-ex/0602035})

\bibitem{dirac}
Dirac P 1928 {\em Proc. R. Soc. (London)\/} {\bf A117, 610 and A118, 351}

\bibitem{snowmass}
Albrecht J {\em et~al.\/} (Intensity Frontier Charged Lepton Working Group)
  2013 {Working Group Report: Charged Leptons} {\em {Proceedings, Community
  Summer Study 2013: Snowmass on the Mississippi (CSS2013): Minneapolis, MN,
  USA, July 29-August 6, 2013}\/} (\textit{Preprint} \eprint{1311.5278})
  \urlprefix\url{https://inspirehep.net/record/1265506/files/arXiv:1311.5278.p%
df}

\bibitem{smqed}
Aoyama T, Hayakawa M, Kinoshita T and Nio M 2012 {\em Phys. Rev. Lett.\/} {\bf
  109}(11) 111807
  \urlprefix\url{http://link.aps.org/doi/10.1103/PhysRevLett.109.111807}

\bibitem{smew}
Gnendiger C, Stockinger D and Stockinger-Kim H 2013 {\em Phys. Rev.\/} {\bf
  D88} 053005 (\textit{Preprint} \eprint{1306.5546})

\bibitem{smhpv1}
Davier M, Hoecker A, Malaescu B and Zhang Z 2011 {\em Eur. Phys. J.\/} {\bf
  C71} 1515 [Erratum: Eur. Phys. J.C72,1874(2012)] (\textit{Preprint}
  \eprint{1010.4180})

\bibitem{smhpv2}
Hagiwara K, Liao R, Martin A~D, Nomura D and Teubner T 2011 {\em J. Phys.\/}
  {\bf G38} 085003 (\textit{Preprint} \eprint{1105.3149})

\bibitem{smhlbl}
Prades J, de~Rafael E and Vainshtein A 2009 {\em Adv. Ser. Direct. High Energy
  Phys.\/} {\bf 20} 303--317 (\textit{Preprint} \eprint{0901.0306})

\bibitem{codata}
Mohr P~J, Taylor B~N and Newell D~B 2008 {\em Rev. Mod. Phys.\/} {\bf 80}(2)
  633--730 \urlprefix\url{http://link.aps.org/doi/10.1103/RevModPhys.80.633}

\bibitem{cern}
Farley F and Picasso E 1990 The cern (g-2) measurements {\em Quantum
  Electrodynamics\/} ed Kinoshita T (World Scientific) p 479

\end{thebibliography}

\end{document}